\RequirePackage{lineno}

\documentclass[aps,prd,twocolumn,superscriptaddress,showpacs]{revtex4}

\usepackage{graphicx}
\usepackage{dcolumn}
\usepackage{bm}
\usepackage{slashed}
\usepackage{subfig}

\RequirePackage{xspace}
\usepackage{relsize}

\def\Kbar  {\kern 0.2em\overline{\kern -0.2em K}{}\xspace}

\def\Bbar    {\kern 0.18em\overline{\kern -0.18em B}{}\xspace}

\def\Qbar    {\kern 0.08em\overline{\kern -0.08em Q}{}\xspace}

\def\invfb   {\ensuremath{\mbox{\,fb}^{-1}}\xspace}
\newcommand{\mev}{\ensuremath{\mathrm{\,Me\kern -0.1em V}}\xspace}
\newcommand{\mevc}{\ensuremath{{\mathrm{\,Me\kern -0.1em V\!/}c}}\xspace}
\newcommand{\mevcc}{\ensuremath{{\mathrm{\,Me\kern -0.1em V\!/}c^2}}\xspace}
\newcommand{\gev}{\ensuremath{\mathrm{\,Ge\kern -0.1em V}}\xspace}
\newcommand{\gevc}{\ensuremath{{\mathrm{\,Ge\kern -0.1em V\!/}c}}\xspace}
\newcommand{\gevcnospace}{\ensuremath{{\mathrm{\,Ge\kern -0.1em V\!/}c}}}
\newcommand{\gevcc}{\ensuremath{{\mathrm{\,Ge\kern -0.1em V\!/}c^2}}\xspace}
\newcommand{\VEt}{\mbox{$\vec{E}_T$}}
\newcommand{\Vpt}{\mbox{$\vec{p}_T$}}

\newcommand{\Et}{\mbox{$E_T$}}
\newcommand{\Pt}{\mbox{$p_T$}}

\newcommand{\met}{\mbox{$\protect \raisebox{0.3ex}{$\not$}\Et$}}
\newcommand{\Vmet}{\mbox{$\protect \raisebox{0.3ex}{$\not$}\VEt$}}
\newcommand{\mpt}{\mbox{$\protect \raisebox{0.3ex}{$\not$}\Pt$}}
\newcommand{\Vmpt}{\mbox{$\protect \raisebox{0.3ex}{$\not$}\Vpt$}}

\begin{document}
\bibliographystyle{apsrev}

\title{{Search for new physics in \boldmath{$t \bar t + \met \to b\bar b q \bar q q \bar q + \met$} final state in $p \bar p$ collisions at $\sqrt{s}=1.96$\,TeV}}

\affiliation{Institute of Physics, Academia Sinica, Taipei, Taiwan 11529, Republic of China} 
\affiliation{Argonne National Laboratory, Argonne, Illinois 60439, USA} 
\affiliation{University of Athens, 157 71 Athens, Greece} 
\affiliation{Institut de Fisica d'Altes Energies, ICREA, Universitat Autonoma de Barcelona, E-08193, Bellaterra (Barcelona), Spain} 
\affiliation{Baylor University, Waco, Texas 76798, USA} 
\affiliation{Istituto Nazionale di Fisica Nucleare Bologna, $^{aa}$University of Bologna, I-40127 Bologna, Italy} 
\affiliation{University of California, Davis, Davis, California 95616, USA} 
\affiliation{University of California, Irvine, Irvine, California 92697, USA} 
\affiliation{University of California, Los Angeles, Los Angeles, California 90024, USA} 
\affiliation{Instituto de Fisica de Cantabria, CSIC-University of Cantabria, 39005 Santander, Spain} 
\affiliation{Carnegie Mellon University, Pittsburgh, Pennsylvania 15213, USA} 
\affiliation{Enrico Fermi Institute, University of Chicago, Chicago, Illinois 60637, USA}
\affiliation{Comenius University, 842 48 Bratislava, Slovakia; Institute of Experimental Physics, 040 01 Kosice, Slovakia} 
\affiliation{Joint Institute for Nuclear Research, RU-141980 Dubna, Russia} 
\affiliation{Duke University, Durham, North Carolina 27708, USA} 
\affiliation{Fermi National Accelerator Laboratory, Batavia, Illinois 60510, USA} 
\affiliation{University of Florida, Gainesville, Florida 32611, USA} 
\affiliation{Laboratori Nazionali di Frascati, Istituto Nazionale di Fisica Nucleare, I-00044 Frascati, Italy} 
\affiliation{University of Geneva, CH-1211 Geneva 4, Switzerland} 
\affiliation{Glasgow University, Glasgow G12 8QQ, United Kingdom} 
\affiliation{Harvard University, Cambridge, Massachusetts 02138, USA} 
\affiliation{Division of High Energy Physics, Department of Physics, University of Helsinki and Helsinki Institute of Physics, FIN-00014, Helsinki, Finland} 
\affiliation{University of Illinois, Urbana, Illinois 61801, USA} 
\affiliation{The Johns Hopkins University, Baltimore, Maryland 21218, USA} 
\affiliation{Institut f\"{u}r Experimentelle Kernphysik, Karlsruhe Institute of Technology, D-76131 Karlsruhe, Germany} 
\affiliation{Center for High Energy Physics: Kyungpook National University, Daegu 702-701, Korea; Seoul National University, Seoul 151-742, Korea; Sungkyunkwan University, Suwon 440-746, Korea; Korea Institute of Science and Technology Information, Daejeon 305-806, Korea; Chonnam National University, Gwangju 500-757, Korea; Chonbuk National University, Jeonju 561-756, Korea} 
\affiliation{Ernest Orlando Lawrence Berkeley National Laboratory, Berkeley, California 94720, USA} 
\affiliation{University of Liverpool, Liverpool L69 7ZE, United Kingdom} 
\affiliation{University College London, London WC1E 6BT, United Kingdom} 
\affiliation{Centro de Investigaciones Energeticas Medioambientales y Tecnologicas, E-28040 Madrid, Spain} 
\affiliation{Massachusetts Institute of Technology, Cambridge, Massachusetts 02139, USA} 
\affiliation{Institute of Particle Physics: McGill University, Montr\'{e}al, Qu\'{e}bec, Canada H3A~2T8; Simon Fraser University, Burnaby, British Columbia, Canada V5A~1S6; University of Toronto, Toronto, Ontario, Canada M5S~1A7; and TRIUMF, Vancouver, British Columbia, Canada V6T~2A3} 
\affiliation{University of Michigan, Ann Arbor, Michigan 48109, USA} 
\affiliation{Michigan State University, East Lansing, Michigan 48824, USA}
\affiliation{Institution for Theoretical and Experimental Physics, ITEP, Moscow 117259, Russia}
\affiliation{University of New Mexico, Albuquerque, New Mexico 87131, USA} 
\affiliation{Northwestern University, Evanston, Illinois 60208, USA} 
\affiliation{The Ohio State University, Columbus, Ohio 43210, USA} 
\affiliation{Okayama University, Okayama 700-8530, Japan} 
\affiliation{Osaka City University, Osaka 588, Japan} 
\affiliation{University of Oxford, Oxford OX1 3RH, United Kingdom} 
\affiliation{Istituto Nazionale di Fisica Nucleare, Sezione di Padova-Trento, $^{bb}$University of Padova, I-35131 Padova, Italy} 
\affiliation{LPNHE, Universite Pierre et Marie Curie/IN2P3-CNRS, UMR7585, Paris, F-75252 France} 
\affiliation{University of Pennsylvania, Philadelphia, Pennsylvania 19104, USA}
\affiliation{Istituto Nazionale di Fisica Nucleare Pisa, $^{cc}$University of Pisa, $^{dd}$University of Siena and $^{ee}$Scuola Normale Superiore, I-56127 Pisa, Italy} 
\affiliation{University of Pittsburgh, Pittsburgh, Pennsylvania 15260, USA} 
\affiliation{Purdue University, West Lafayette, Indiana 47907, USA} 
\affiliation{University of Rochester, Rochester, New York 14627, USA} 
\affiliation{The Rockefeller University, New York, New York 10065, USA} 
\affiliation{Istituto Nazionale di Fisica Nucleare, Sezione di Roma 1, $^{ff}$Sapienza Universit\`{a} di Roma, I-00185 Roma, Italy} 

\affiliation{Rutgers University, Piscataway, New Jersey 08855, USA} 
\affiliation{Texas A\&M University, College Station, Texas 77843, USA} 
\affiliation{Istituto Nazionale di Fisica Nucleare Trieste/Udine, I-34100 Trieste, $^{gg}$University of Trieste/Udine, I-33100 Udine, Italy} 
\affiliation{University of Tsukuba, Tsukuba, Ibaraki 305, Japan} 
\affiliation{Tufts University, Medford, Massachusetts 02155, USA} 
\affiliation{University of Virginia, Charlottesville, VA  22906, USA}
\affiliation{Waseda University, Tokyo 169, Japan} 
\affiliation{Wayne State University, Detroit, Michigan 48201, USA} 
\affiliation{University of Wisconsin, Madison, Wisconsin 53706, USA} 
\affiliation{Yale University, New Haven, Connecticut 06520, USA} 
\author{T.~Aaltonen}
\affiliation{Division of High Energy Physics, Department of Physics, University of Helsinki and Helsinki Institute of Physics, FIN-00014, Helsinki, Finland}
\author{B.~\'{A}lvarez~Gonz\'{a}lez$^w$}
\affiliation{Instituto de Fisica de Cantabria, CSIC-University of Cantabria, 39005 Santander, Spain}
\author{S.~Amerio}
\affiliation{Istituto Nazionale di Fisica Nucleare, Sezione di Padova-Trento, $^{bb}$University of Padova, I-35131 Padova, Italy} 

\author{D.~Amidei}
\affiliation{University of Michigan, Ann Arbor, Michigan 48109, USA}
\author{A.~Anastassov}
\affiliation{Northwestern University, Evanston, Illinois 60208, USA}
\author{A.~Annovi}
\affiliation{Laboratori Nazionali di Frascati, Istituto Nazionale di Fisica Nucleare, I-00044 Frascati, Italy}
\author{J.~Antos}
\affiliation{Comenius University, 842 48 Bratislava, Slovakia; Institute of Experimental Physics, 040 01 Kosice, Slovakia}
\author{G.~Apollinari}
\affiliation{Fermi National Accelerator Laboratory, Batavia, Illinois 60510, USA}
\author{J.A.~Appel}
\affiliation{Fermi National Accelerator Laboratory, Batavia, Illinois 60510, USA}
\author{A.~Apresyan}
\affiliation{Purdue University, West Lafayette, Indiana 47907, USA}
% ...
\collaboration{CDF Collaboration\footnote{With visitors from $^a$University of Massachusetts Amherst, Amherst, Massachusetts 01003,
$^b$Istituto Nazionale di Fisica Nucleare, Sezione di Cagliari, 09042 Monserrato (Cagliari), Italy,
$^c$University of California Irvine, Irvine, CA  92697, 
$^d$University of California Santa Barbara, Santa Barbara, CA 93106
$^e$University of California Santa Cruz, Santa Cruz, CA  95064,
$^f$CERN,CH-1211 Geneva, Switzerland,
$^g$Cornell University, Ithaca, NY  14853, 
$^h$University of Cyprus, Nicosia CY-1678, Cyprus, 
$^i$University College Dublin, Dublin 4, Ireland,
$^j$University of Fukui, Fukui City, Fukui Prefecture, Japan 910-0017,
$^k$Universidad Iberoamericana, Mexico D.F., Mexico,
$^l$Iowa State University, Ames, IA  50011,
$^m$University of Iowa, Iowa City, IA  52242,
$^n$Kinki University, Higashi-Osaka City, Japan 577-8502,
$^o$Kansas State University, Manhattan, KS 66506,
$^p$University of Manchester, Manchester M13 9PL, England,
$^q$Queen Mary, University of London, London, E1 4NS, England,
$^r$University of Melbourne, Victoria 3010, Australia,
$^s$Muons, Inc., Batavia, IL 60510,
$^t$Nagasaki Institute of Applied Science, Nagasaki, Japan, 
$^u$National Research Nuclear University, Moscow, Russia,
$^v$University of Notre Dame, Notre Dame, IN 46556,
$^w$Universidad de Oviedo, E-33007 Oviedo, Spain, 
$^x$Texas Tech University, Lubbock, TX  79609, 
$^y$Universidad Tecnica Federico Santa Maria, 110v Valparaiso, Chile,
$^z$Yarmouk University, Irbid 211-63, Jordan,
$^{hh}$On leave from J.~Stefan Institute, Ljubljana, Slovenia, 
}}
\noaffiliation

\begin{abstract}
We present a search for a new particle $T^\prime$ decaying to a top quark via $T^{\prime} \rightarrow t + X$, where $X$ goes undetected. We use a data sample corresponding to 5.7 fb$^{-1}$ of integrated luminosity of $p\bar{p}$ collisions with $\sqrt{s} = 1.96$ TeV, collected at Fermilab Tevatron by the CDF II detector. Our search for pair production of $T^\prime$ is focused on the hadronic decay channel, $p\bar{p} \rightarrow T^\prime\bar{T^\prime} \rightarrow t\bar{t} +X\bar X \rightarrow  bq\bar q \,  \bar{b}q \bar{q} \, + X\bar X$. We interpret our results in terms of a model where $T^\prime$ is an exotic fourth generation quark and $X$ is a dark matter particle. The data are consistent with standard model expectations. We set a limit on the generic production of $T^\prime\bar{T^\prime} \rightarrow t\bar{t} \, + \, X\bar X$, excluding the fourth generation exotic quarks $T^\prime$ at 95\% confidence level up to $m_{T^\prime}=400$ GeV$/c^{2}$ for $m_X\leq 70$ GeV$/c^2$.

\end{abstract}



\maketitle

\date{\today}

There are many hints, from astronomical observations and  cosmological theories, for the existence of dark matter (DM) particles, which must be long-lived on cosmological time scale \cite{Feng:2010gw}. The long lifetime of DM can be explained by the conservation of a charge of a new symmetry.
Direct-detection experiments based on ultra-low noise devices have recently obtained interesting results. The DAMA/LIBRA Collaboration \cite{Bernabei:2008yi}, searching for annual  modulation in the interaction rate due to the earth motion through DM galactic halo, has claimed a $\simeq 9\sigma$ observation of DM. The CoGeNT Collaboration has also reported evidence of DM\,\cite{Aalseth:2011wp}.
If confirmed, these results would imply, unlike astronomical observations, that DM interactions with standard model (SM) particles are not only gravitational.  DM interactions with SM particles could be allowed by weak interactions, or by connector particles carrying both dark and SM charges, so that they could be produced in colliders. The second hypothesis is favored in the case that DM particles have mass of a few GeV/$c^2$, as DAMA and CoGeNT results seems to indicate. In a recent model \cite{Alwall:2010jc} the role of a connector particle is played by an exotic fourth generation $T^\prime$ quark, which is supposed to decay to a top quark and dark matter, $T^\prime\rightarrow t + X$. The pair production of such exotic quarks and their subsequent decay has a collider signature consisting  of top quark pairs ($t\bar t$) and missing transverse energy ($\met$)~\cite{met_def} due to	the invisible dark matter particles. These types of signals  are of great interest as they appear also in other models containing DM candidates, such as scalar top quarks production and their decay to top quarks and neutralinos \cite{Ellis:2010kf} or top quarks and gravitinos \cite{Kats:2011it}, and in many other new physics scenarios such as little Higgs \cite{Cheng:2004yc} and models where baryon and lepton numbers represent local gauge symmetries \cite{FileviezPerez:2010gw}.

A first search for the $T^{\prime}\bar{T^\prime} \rightarrow t \bar{t} + X\bar X$ process  has been performed in the semileptonic channel: $t \bar{t} +X\bar X \rightarrow bW \bar{b}W + X\bar X \rightarrow b l\nu \bar{b}  q\bar{q} + X\bar X$ \cite{Aaltonen:2011rr}. This letter reports the first search for such a process in the all-hadronic $t\bar{t}$ decay channel, characterized by a larger branching ratio and a lower physics background rate. 
%
%
Events were recorded by CDF II~\cite{cdf} a general purpose detector designed to study collisions at the Fermilab Tevatron $p\overline{p}$ collider at $\sqrt{s}=1.96$ TeV. The tracking system consists of a cylindrical open-cell drift chamber and silicon microstrip detectors in a 1.4 T magnetic field parallel to the beam axis. Electromagnetic and hadronic calorimeters surrounding the tracking system measure particle energies and drift chambers located outside the calorimeters detect muons. Jets are reconstructed in the calorimeter using the {\sc jetclu}~\cite{jetclu} algorithm with a clustering radius of 0.4 in azimuth-pseudorapidity space~\cite{coordinates}. The detector response for all simulated samples is modeled by a detailed CDF detector simulation.
Production of $T^\prime$ pairs and their subsequent decays to top quark pairs and two dark matter particles would appear as events with missing transverse energy from the two dark matter particles, and six jets from the two $b$ quarks and the hadronic decays of the two $W$ bosons. We model the production  and decay of $T^\prime$ pairs with the {\sc madgraph} Monte Carlo (MC) generator~\cite{madgraph}, and normalize to the next-to-next-to-leading order (NNLO) cross section calculation \cite{Cacciari:2008zb}. Additional radiation, hadronization and showering are described by {\sc pythia} \cite{pythia}.

\begin{figure}[!h]
\begin{center}
\includegraphics[width=0.9\linewidth]{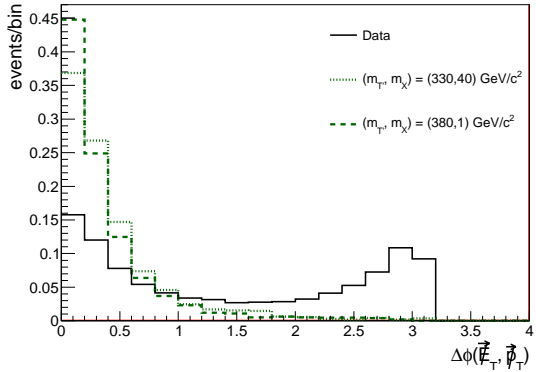}\\
\caption{Distribution of $\Delta\phi (\Vmet, \Vmpt)$ for the preselection data, and two scenarios with different values of $m_{T^\prime}$ and $m_{X}$. All histograms are normalized to unit area.}
\label{fig:DPhiMET_MPT}
\end{center}
\end{figure}

\begin{figure}[!htb]
\begin{center}
\includegraphics[width=0.74 \linewidth]{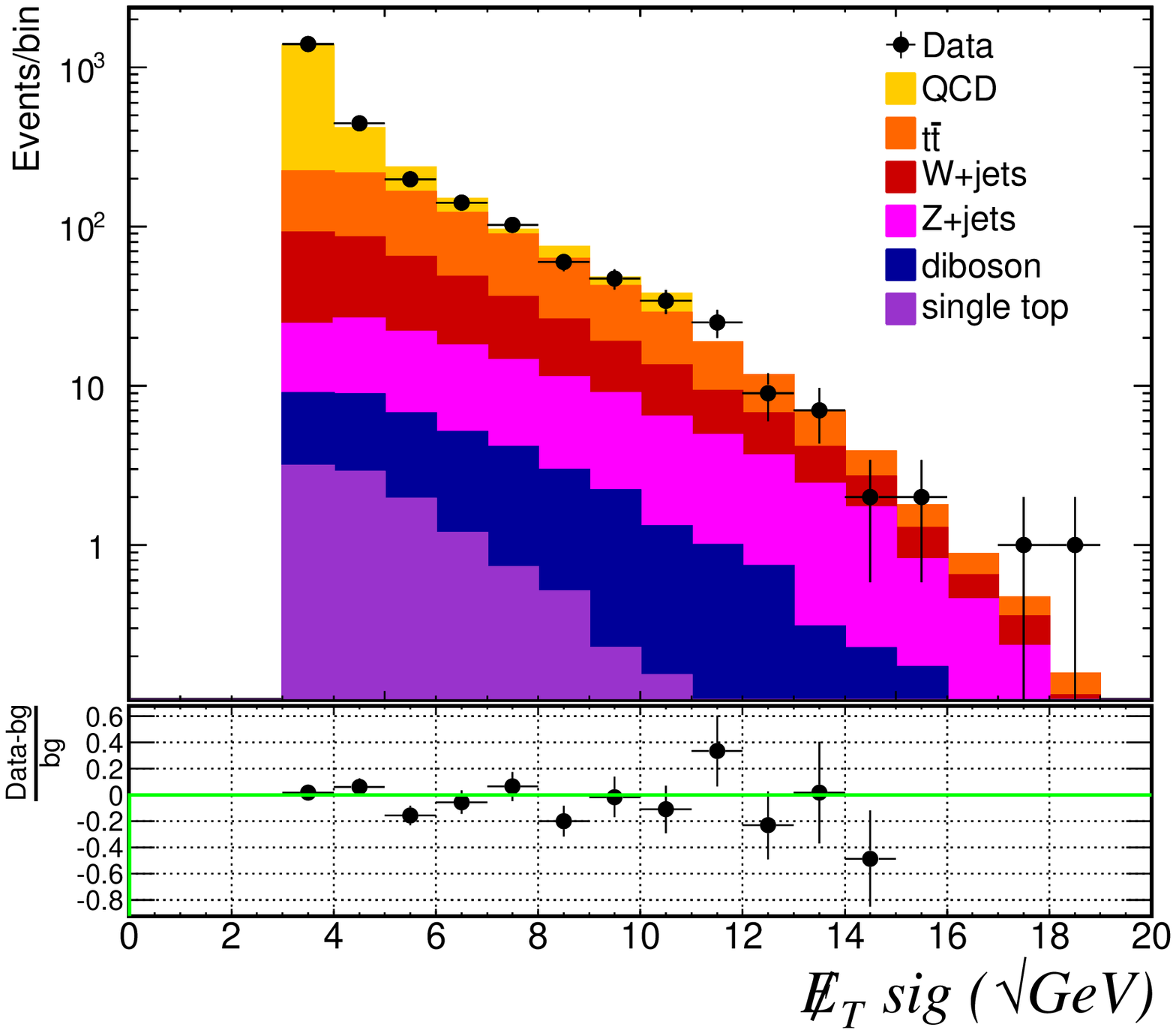}\\
\includegraphics[width=0.74 \linewidth]{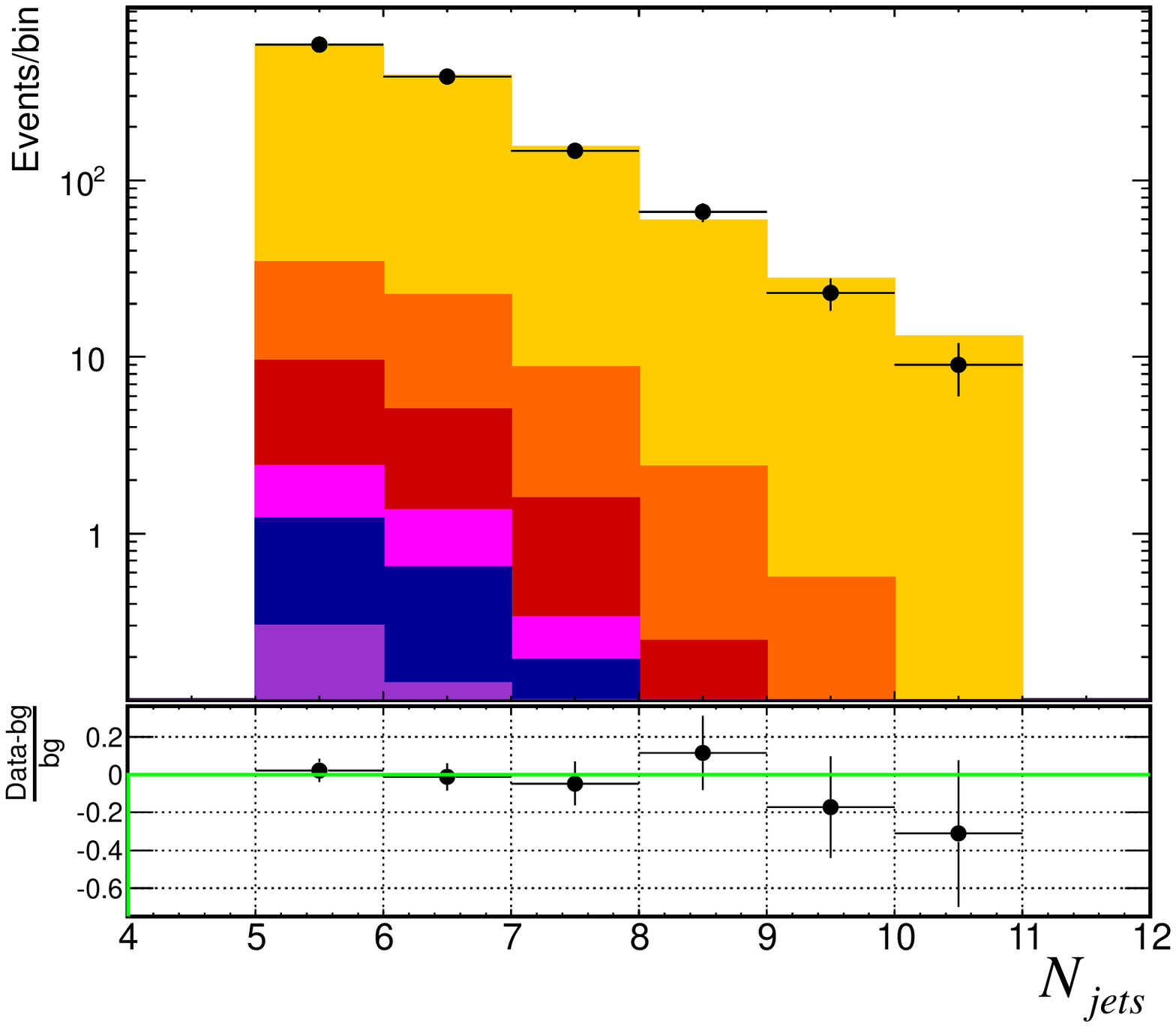}\\
\caption{Top plot shows the  $\met \, sig$ distribution in events with four jets and large $\met$. Bottom plot shows the  ${N}_{jets}$ distribution in event with $5 \leq N_{jets} \leq 10$ and $\met \, sig < 3\,\sqrt{GeV}$.}
\label{fig:ControlReg}
\end{center}
\end{figure}

\begin{figure}[!h]
\begin{center}
\includegraphics[width=1.0\linewidth]{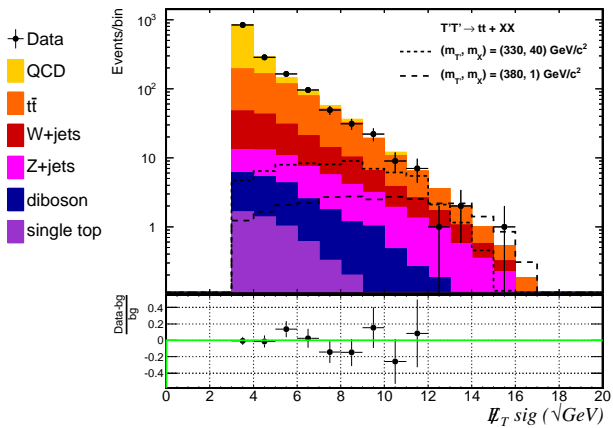}\\
\caption{$\met \, sig$ distributions for the standard model backgrounds, the observed data, and for two scenarios with different values of $m_{T^\prime}$ and $m_{X}$.}
\label{fig:signals}
\end{center}
\end{figure}

\begin{figure}[!h]
\begin{center}
\includegraphics[width=1.0\linewidth]{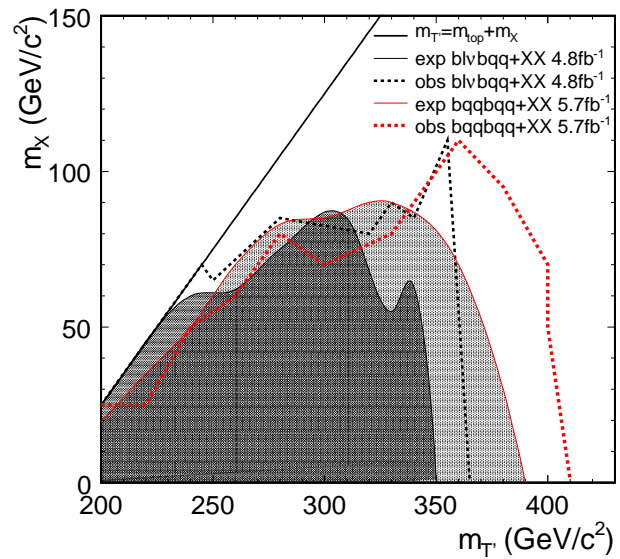}\\
\caption{Expected (exp) and observed (obs) 95\% C.L. exclusion region in the ($m_{T^{\prime}}, m_{X}$) parameters space.}
\label{fig:sigmadataexclusion}
\end{center}
\end{figure}

We use a data sample corresponding to an integrated luminosity of $p\bar{p}$ collisions of 5.7 \invfb, collected by requiring  $\met>$ 50 GeV  and two or more jets with transverse energy $E_T \geq$ 30 GeV and $|\eta| \leq 2.4$ .
We then require $5 \leq N_{jets} \leq 10$, where $N_{jets}$ is the number of jets, and where all jets satisfy the requirement $|\eta| \leq 2.4$. We also require the transverse energy  $E_{\mathrm{T}}$ of the subleading jets, $J_i$, to be greater than 20 GeV for $(i = 3,4,5)$ and 15 GeV for $(i > 5)$. We veto events with at least one isolated electron or muon to suppress events with semileptonic $t \bar{t}$ decay. We refer to this sample as the preselection sample. 
At this stage of the event selection, multijet QCD background where $\met$ arises from jet energy mismeasurement accounts for more than 95\% of the expected backgrounds. 
The second dominant background is $t \bar{t}$ production. We model this process using {\sc pythia} with $m_t = 172.5$ GeV$/c^2$ \cite{topmass}, normalized to the next-to-leading (NLO) order cross section~\cite{nloa}.
Associated production of $W/Z$ boson and jets is also a significant background source. Samples of simulated $W/Z$+jets events with light- and heavy-flavor jets are generated using the {\sc alpgen}~\cite{alpgen} MC generator, interfaced with the parton shower model of {\sc pythia}. A matching scheme is applied to avoid double-counting of partonic event configurations\,\cite{Mangano:2006rw}. The $W/Z$+jets samples are normalized to the measured $W$ and $Z$  cross section ~\cite{ewknlo}.
Diboson and single top production are modeled using respectively {\sc pythia} and {\sc madgraph}, and normalized to NLO cross sections~\cite{ewknlo,singletops,singletopt,Campbell:2009ss}. 
Because of the large production rate for QCD multijet events at a hadron collider and the statistics needed in order to describe this process adequately in an analysis looking for a very small signal, the Monte Carlo simulation of QCD multijet events is prohibitive. More importantly, the systematic uncertainties associated with the Monte Carlo simulation of QCD jet production are large. For these reasons, we estimate the QCD background solely from data.
Similarly to $\Vmet$, it is possible to define a missing transverse momentum $\Vmpt$ using the spectrometer, as the negative vector sum of the charged particles momenta. $\Vmet$ and $\Vmpt$ are correlated in magnitude and direction in events with undetected particles. In QCD multijet events $\Vmet$ originates from the mismeasurement of a jet energy in the calorimeter, while  $\Vmpt$ depends on fluctuations in the number of charged particles in a jet, so they are usually aligned or anti-aligned in dijet-like events like energetic QCD multijet events, as is shown in Fig.~\ref{fig:DPhiMET_MPT}. QCD multijet events in which  $\Vmet$ and $\Vmpt$ are aligned or anti-aligned have the same kinematic characteristics, as we have verified studying QCD multijet samples with 2 and 3 jets\,\cite{Aaltonen:2010fs}.
%
%
We reject events with $\Delta\phi (\Vmet, \Vmpt) > \pi/2$, and use them to model QCD multijet events in the signal region $\Delta\phi (\Vmet, \Vmpt) < \pi/2$. 
To further suppress the QCD multijet background, we require the azimuthal distance between the directions of $\Vmet$ and subleading jets, $\Delta\phi (\Vmet , \vec{J}_i)$, to be greater than 0.4  for $i = 1,2,3$ and 0.2 for $i = 4,5$. We also require $\mpt > 20$ GeV and $\met \, sig > 3 GeV^{1/2}$, where  $\met \, sig$ is defined as the $\met$ divided by the square root of the total energy collected in the calorimeter. Finally, we require $\sum_{jets} E_T^i > 220$ GeV to remove soft QCD events. All these cuts have been chosen to optimize the $S/\surd(S+B)$ figure of merit, where S and B are respectively the expected numbers of signal and background events. Table~\ref{Table1} shows the expected number of events in the signal region for SM backgrounds and for several signal hypothesis.
\begin{center}
\begin{table}[!h]
\caption{Number of expected signal events for five benchmark scenarios compared to data and expected SM backgrounds. 
}
\begin{tabular} {c c} \hline\hline
$T^{\prime} \bar T^{\prime}\rightarrow t \bar t X\bar X (hadronic)$ [GeV$/c^{2}$] &   Events \\  \hline
$m_{T^{\prime}},m_{X}=260,80$            &     $88.5 \pm 11.9$ \\ 
$m_{T^{\prime}},m_{X}=330,100$           &     $66.4 \pm 8.9$  \\ 
$m_{T^{\prime}},m_{X}=360,100$           &     $39.7 \pm 5.3$ \\  
$m_{T^{\prime}},m_{X}=380,1$             &     $27.3 \pm 3.7$\\
$m_{T^{\prime}},m_{X}=400,1$             &     $17.5 \pm 2.3$\\ 
QCD                                      &     $745.4\pm 124.3$  \\
t\={t}                                   &     $498.2 \pm 66.8$ \\ 
W+jets                                   &     $119.7 \pm 48.4$ \\ 
Z+jets                                   &     $39.4 \pm 15.9$   \\ 
Diboson                                  &     $17.9 \pm 2.2$  \\ 
Single top                               &     $5.3 \pm 0.8$   \\ \hline
Total Background                         &     $1423 \pm 150$  \\ \hline
Data                                     &          1507  \\ \hline\hline
\end{tabular}
\label{Table1}
\end{table}
\end{center}
Inverting one of the event selection cuts, keeping others unchanged, allows us to define a signal-depleted control region. We use $\met \, sig < 3 \, GeV^{1/2}$, $N_{jets} = 4$ and $\mpt < 20$ GeV control regions to validate the overall background modeling. The normalization factor of the QCD background is given by the average ratio of QCD events that pass the $\Delta\phi (\Vmet, \Vmpt) < \pi/2$ cut to QCD events that fail the cut in these three control regions. Figure~\ref{fig:ControlReg} shows good agreement of background modeling with data in these regions.
We consider several sources of systematic uncertainties. The dominant components are the uncertainties on the QCD normalization factor, the jet energy scale (JES)~\cite{jes} and the theoretical cross sections. We also take in account the differences of $t \bar{t}$ predicted rates using different hadron fragmentation models in the {\sc herwig}\,\cite{herwig} Monte Carlo, and varying initial/final state radiation and  color reconnection effects\,\cite{CR}.
The variation of the JES was found to change significantly the $\met \, sig$ distribution in addition to its normalization, and its variation is thus taken into account.
%
%
Figure~\ref{fig:signals} shows the $\met \, sig$ distribution for expected signal events and SM backgrounds. 
The signal is expected to contribute significantly in the high tail of the $\met \, sig$ distribution. 
There is no evidence for the presence of $T^{\prime} \rightarrow t + X$ events in the data. We calculate 95\% C.L. upper limits on the $T^{\prime} \rightarrow t + X$ cross section, by performing a binned maximum-likelihood fit on the $\met \, sig$ distribution. The limits are calculated using a Bayesian likelihood method with a flat prior for the signal cross-section, integrating over Gaussian priors for the systematic uncertainties.
The results are shown in Table~\,\ref{tab:Limits}. We convert the observed upper limits on the pair-production cross sections to an exclusion curve in mass parameters space $(m_{T^\prime}, m_X)$. As shown in Fig.~\ref{fig:sigmadataexclusion}, a significant enhancement in sensitivity is obtained when comparing to the previous analysis in semi-leptonic channel.
\begin{table}[!h]
  \begin{center}
    \caption{Expected 95\% C.L. upper limits on $T^\prime T^\prime$ production, where the uncertainty range covers 68\% of the pseudoexperiments, and observed exclusion limits for representative signal points.}
    \begin{tabular}{c c c}
      \hline\hline
$(mT^\prime,mX)$ $GeV/c^2$   &  ${\sigma}_{exp. 95\%C.L. excl.}(pb)$  & ${\sigma}_{obs. 95\%C.L. excl.}(pb)$ \\ \hline
(200,40)       &           2.02 $\pm$ 0.65                  &                       1.90       \\ 
(220,40)      &            2.14 $\pm$ 0.75                  &                       3.00       \\
(260,1)       &            0.23 $\pm$ 0.08                  &                       0.18       \\
(280,1)       &            0.15 $\pm$ 0.05                  &                       0.12       \\
(280,40)      &            0.18 $\pm$ 0.07                  &                       0.15      \\
(300,1)      &             0.09 $\pm$ 0.03                  &                       0.09 \\  
(300,80)      &            0.20 $\pm$ 0.06                  &                       0.16 \\  
(300,100)     &            0.29 $\pm$ 0.09                  &                       0.38 \\  
(330,1)     &              0.05 $\pm$ 0.02                  &                       0.03 \\ 
(330,100)     &            0.13 $\pm$ 0.04                  &                       0.18 \\ 
(360,1)     &              0.03 $\pm$ 0.01                  &                       0.02 \\ 
(360,100)     &            0.06 $\pm$ 0.02                  &                       0.04 \\ 
(380,100)     &            0.06 $\pm$ 0.02                  &                       0.05 \\ 
(400,1)     &              0.023 $\pm$ 0.008                &                       0.016\\  \hline
\hline
    \end{tabular}
    \label{tab:Limits}
  \end{center}
\end{table}

In conclusion, we performed the first search for new physics in the $t \bar t + \met \to b\bar b q \bar q q \bar q + \met$ final state. Data is consistent with the background-only hypothesis, and we thus set 95\%~C.L. upper limit on the production cross section for fermionic $T^\prime$ pairs decaying to top quarks and dark matter candidates $X$, increasing the existing mass exclusion range up to $m_{T^\prime}=400$ GeV$/c^{2}$, for $m_X\le 70$ GeV$/c^2$. Finally, this study shows that the $b\bar b q \bar q q \bar q + \met$ final state is the most sensitive to the generic production of top quarks plus dark matter candidates, and thus the most promising to probe the supersymmetric $\tilde t \to t + \chi/g$ scenarios at the LHC.

We thank Johan Alwall and Matteo Cacciari for the useful discussions. We also thank the Fermilab staff and the technical staffs of the participating institutions for their vital contributions. This work was supported by the U.S. Department of Energy and National Science Foundation; the Italian Istituto Nazionale di Fisica Nucleare; the Ministry of Education, Culture, Sports, Science and Technology of Japan; the Natural Sciences and Engineering Research Council of Canada; the National Science Council of the Republic of China; the Swiss National Science Foundation; the A.P. Sloan Foundation; the Bundesministerium f\"ur Bildung und Forschung, Germany; the Korean World Class University Program, the National Research Foundation of Korea; the Science and Technology Facilities Council and the Royal Society, UK; the Institut National de Physique Nucleaire et Physique des Particules/CNRS; the Russian Foundation for Basic Research; the Ministerio de Ciencia e Innovaci\'{o}n, and Programa Consolider-Ingenio 2010, Spain; the Slovak R\&D Agency; the Academy of Finland; and the Australian Research Council (ARC).


\begin{thebibliography}{99}
\bibitem{met_def} 
Missing transverse energy, $\met$, is defined as the magnitude of the vector $\Vmet = -\sum_i E_T^i \vec n_i$ where $E_T^i$ are the magnitudes of transverse energy contained in each calorimeter tower $i$, and $\vec n_i$ is the unit vector from the interaction vertex to the tower in the transverse ($x,y$) plane.
\bibitem{coordinates}
CDF uses a cylindrical coordinate system with
  the $z$ axis along the proton beam axis. Pseudorapidity is $\eta
  \equiv -\ln(\tan (\theta/2))$, where $\theta$ is the polar angle
  relative to the proton beam direction, and $\phi$ is the azimuthal angle  while $p_T = |p|\ \sin\theta$, $E_T = E\ \sin\theta$.
\end{thebibliography}

\end{document}